\documentclass{PoS}

\usepackage{color}
\usepackage{lineno}

\title{Search for Gamma-ray Line Signatures with H.E.S.S.}

\ShortTitle{Search for Gamma-ray Line Signatures with H.E.S.S.}

\author{
M. Kieffer$^{1}$, 
K. D. Mor\aa$^{2}$,
J. Conrad$^{2}$,
C. Farnier$^{2}$,
A. Jacholkowska$^{1}$,
J. Veh$^{3}$,
A. Viana$^{4}$

for the H.E.S.S. Collaboration. 
\\ \\ \\
\llap{$^1$} LPNHE, Universit\'{e} Paris VI et Paris VII, CNRS/IN2P3, 4 Place Jussieu, F-75252, Paris Cedex 5, France. \\
\llap{$^2$} OKC, Physics Department, Stockholm University, AlbaNova SE-10691 Stockholm, Sweden. \\
\llap{$^3$} Universit\"at Erlangen-N\"urnberg, Physikalisches Institut, 91058 Erlangen, Germany. \\
\llap{$^4$} Max-Planck-Institut f\"{u}r Kernphysik, P.O. Box 103980, D 69029 Heidelberg, Germany. \\

E-mail: \email{mkieffer@lpnhe.in2p3.fr}

}

\abstract{Many results from astrophysical observations point to a 27\% contribution of non-baryonic dark matter to the mass-energy budget of the universe. Although still elusive, strongly motivated candidates in form of weakly interacting massive particles could explain the nature of dark matter, and their annihilation or decay would give rise to detectable signatures in gamma-rays. In 2012, the H.E.S.S. collaboration started taking data with the largest imaging atmospheric Cherenkov telescope in the world which significantly lowered the energy threshold of the already operational four-telescope system. In particular, due to its location and improved performance at low energies, the H.E.S.S. experiment is now in a position to extend the search for dark matter line signals down to the 100 GeV mass range. The sensitivity to line searches with a new full likelihood approach will be discussed and preliminary results from observations with the second phase of H.E.S.S. will be presented.}

\FullConference{The 34th International Cosmic Ray Conference,\\
		30 July- 6 August, 2015\\
		The Hague, The Netherlands}

\begin{document}

\section{Introduction}
Weakly interacting massive particles (WIMPs) are currently among the
best candidates to explain the elusive nature of dark matter
(DM) and are actively searched for by the community (see
\cite{Bertone:2004pz} for a review). In particular, indirect detection
using gamma-rays is considered as one of the most promising avenues as
it allows the precise determination of the site of their production, and
high-energy gamma-ray experiments have now the capability to
reconstruct accurately the observed spectra.
The gamma-ray spectrum arising from the annihilation of WIMPs is
characterized by two main components: a continuum of gamma-rays up to
the WIMP mass and possible gamma-ray lines.
The detection of such a line would be a smoking gun signal for the
existence of WIMPs since no standard astrophysical process can mimic
a line signature. Its search is therefore facilitated and much more robust than the
search for the continuum, in particular for regions of the sky with
strong astrophysical gamma-ray emissions \cite{bib:Conrad}.
\\ \\
The H.E.S.S. collaboration reported on the search for line signatures
based on observations of the Galactic centre performed with
H.E.S.S. I, covering a WIMP mass range between 500 GeV and 20 TeV
\cite{hess_line}. The commissioning of the CT5 telescope lowered the energy threshold down to $\sim$80 GeV
at the analysis level for zenith observations. The new
H.E.S.S. II data now allow probing the unexplored mass range below
500~GeV and to achieve a significant overlap with the H.E.S.S. I and
Fermi-LAT results.

\section{Methodology}
\subsection{Analysis strategy}
The general analysis strategy relies on the idea of fitting a line-like signal on top of a measured background energy spectrum in the ON-source region, using a full event-by-event likelihood procedure optimized for DM searches in the Galactic Center region. 
Here no background subtraction was performed therefore preserving the entire potential dark matter signal. Since measured distributions are considered in the fit there is no need for acceptance corrections thus strongly limiting the associated systematic uncertainties.  However, as there is no "universal" background that can be estimated from H.E.S.S. observations due to the difference in night sky background distributions over a given FoV, a careful choice of background control regions was performed. 
\\ \\
For the analysis presented, the region of interest (ROI) is centred at the best fit position of the 130 GeV line feature reported in Fermi-LAT data (\cite{bib:4},\cite{su_fink}) that is found to be displaced by $-1.5^{\circ}$ longitude with respect to the Galactic Center, called here "Fermi hotspot". The data control regions were defined closely and symmetrically surrounding the Fermi hotspot position to estimate the background fraction in the analysis region. Spectral contributions from cosmic-ray background and astrophysical diffuse emission are supposed to be included. The instrument response functions (IRFs) obtained from gamma-ray Monte-Carlo (MC) were used to derive the expected measured energy spectra for the line signals considering the stereo reconstruction algorithm which infers the direction and energy of gamma-like events from both the signal and the background regions.
\\ \\
Sensitivity estimates are provided by MC simulations in section \ref{MC} for a line scan between 100 GeV and 2 TeV, and preliminary results with 95\% confidence level (CL) limits are given in section \ref{FermiHS} on a sub-sample of the total H.E.S.S.II dataset focusing on the particular case of the 130 GeV line feature.

\subsection{Full event-by-event likelihood}
The full likelihood function is composed of a Poisson normalization term (based on the total number of events in the signal and background regions) and a spectral term related to the expected spectral contribution of the signal and the background component in the analysis region (ON-source region). A description of the method applied in this study is given in \cite{bib:Aleksic}. The number of signal ($N_{signal}$) and background ($N_{bckg}$) events in the analysis region are free parameters of the model, while additional information on the signal and background spectral shape is included in the fit. Usually, the extracted parameters are the line signal fraction $\eta = \frac{N_{signal}}{N_{signal}+N_{bckg}}$ which represents the relative contribution of the signal in the analysis region, and the line energy position $E_{line}$. Here results on $\eta$ are presented. The full likelihood formula is:
\begin{eqnarray}
\mathcal{L}(N_{signal},N_{bckg}|N_{ON},N_{OFF},E_{i}) = &\frac{(N_{signal}+N_{bckg})^{N_{ON}}}{N_{ON}!}e^{-(N_{signal}+N_{bckg})} \times \frac{(\alpha N_{bckg})^{N_{OFF}}}{N_{OFF}!}e^{-\alpha N_{bckg}} \nonumber \\ &\times \prod \limits_{i=1}^{N_{ON}} \Big( \eta \times PDF_{signal} (E_{i}) + (1-\eta) \times PDF_{bckg} (E_{i}) \Big)
\label{eq:fulllikelihoodformula}
\end{eqnarray}
where $N_{ON}$ and $N_{OFF}$ are the measured number of events in the signal and background regions, $\alpha$ the exposure ratio between background and signal regions, and $E_{i}$ represent the individual energies of the events in the analysis region. $PDF_{signal}$ and $PDF_{bckg}$ are the probability density functions for the signal and background components that refer to measured energy spectra i.e. photon energies smeared by the response functions of the instrument. $PDF_{signal}$ is a function of the line energy  $E_{line}$.

\subsection{ROI optimization}
An alternative binned likelihood method was developed to perform the ROI optimization. The binned ON-OFF excess is fitted to an IRF-convolved power-law in energy side-bands above and below the line energy region. The excess due to diffuse emission in the line energy region is estimated from the fit. Given the expected and observed number of events in the line energy region, the averaged upper limit is computed using Rolke et. al \cite{bib:Rolke}. \\
Assuming the peak of the DM density profile at the center of the ROI, the size of the signal region is chosen to maximise the sensitivity determined with the binned likelihood method. The optimization was done with a 130 GeV line signal following the annihilation cross-section given in \cite{bib:4}. Diffuse gamma-ray emission is estimated using the H.E.S.S. Galactic Plane Survey \cite{bib:3}, extrapolated to lower energies using a power law of index 2.3, while background due to protons and electrons is estimated using OFF-data from control samples.
An optimal signal region size of $0.4^{\circ}$ is found, corresponding to a solid angle $\Delta \Omega = 1.531*10^{-4} sr$.

\newpage

\section{Sensitivity Estimates}
\label{MC}
\begin{figure}[ht]
\centering
\begin{minipage}[b]{0.47\linewidth}
\includegraphics[width=1\textwidth]{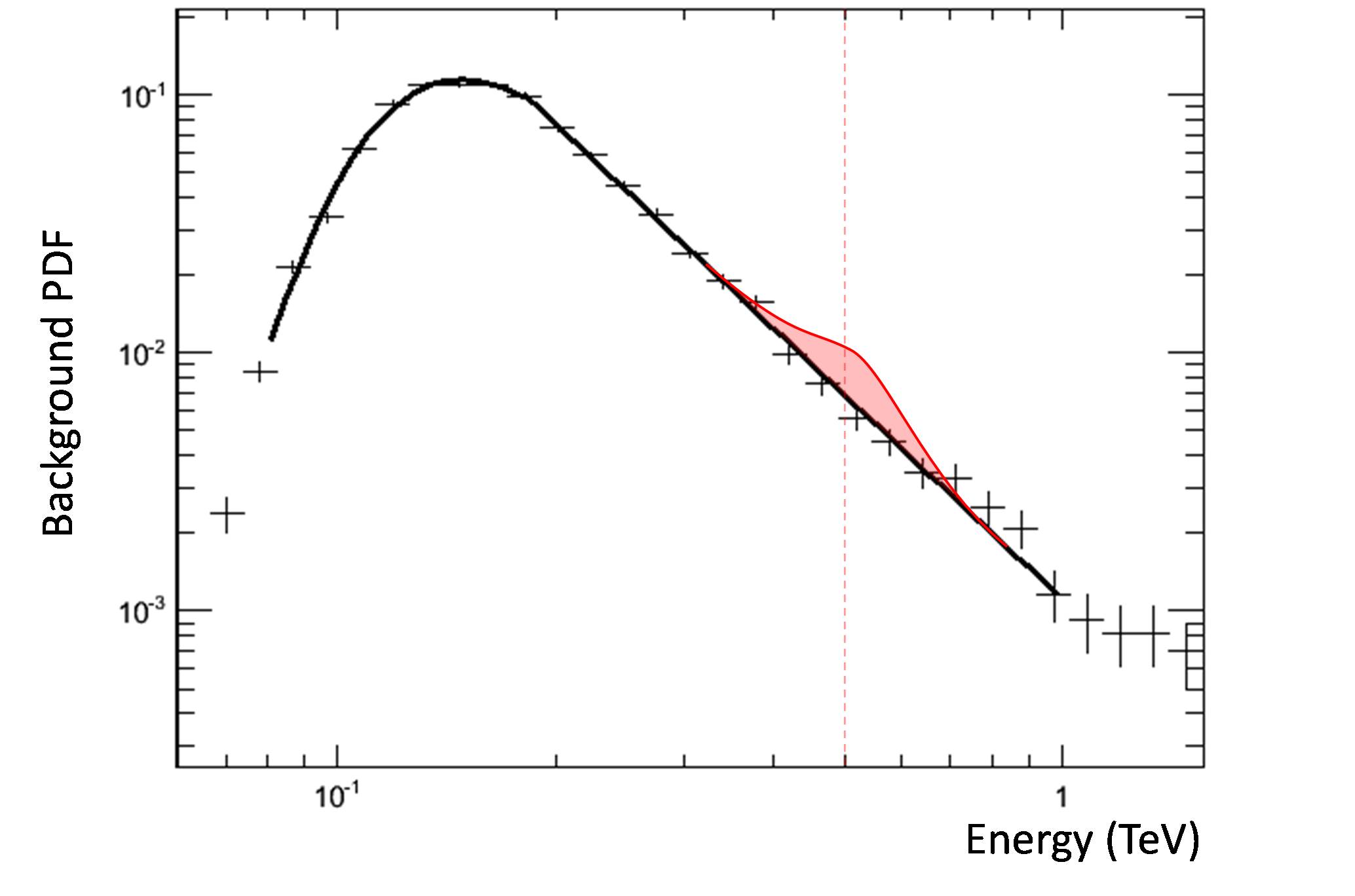}
\caption{Background PDF for the full likelihood method. The PDF is generated from
background control regions around the Fermi hotspot (see results in section \protect\ref{FermiHS}). The line signal PDF at 500 GeV is also represented in red for $\eta=0.05$.}
\label{fig:background_pdf}
\end{minipage}
\quad
\begin{minipage}[b]{0.47\linewidth}
\includegraphics[width=1\textwidth]{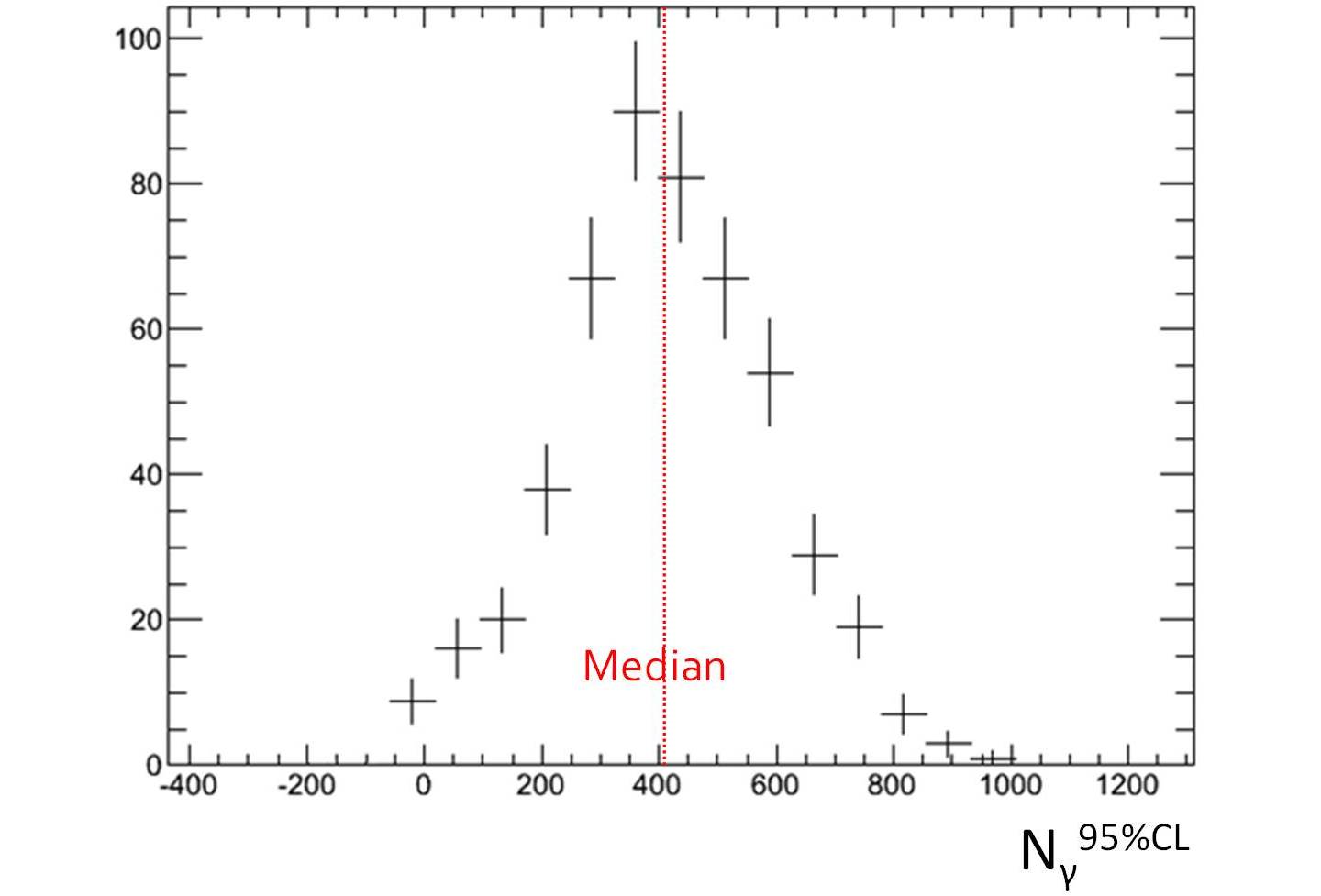}
\caption{Distribution of the $95\%$ CL limit on the number of excess gamma-line events $N_{\gamma}^{95\% CL}$ in the analysis region using the full likelihood method, here with the example of a 130 GeV line. The limit is taken as the median in the 500 MC datasets.}
\label{fig:ngamma_limits_median}
\end{minipage}
\end{figure}

\begin{figure}[ht]
\centering
\includegraphics[width=0.868\textwidth]{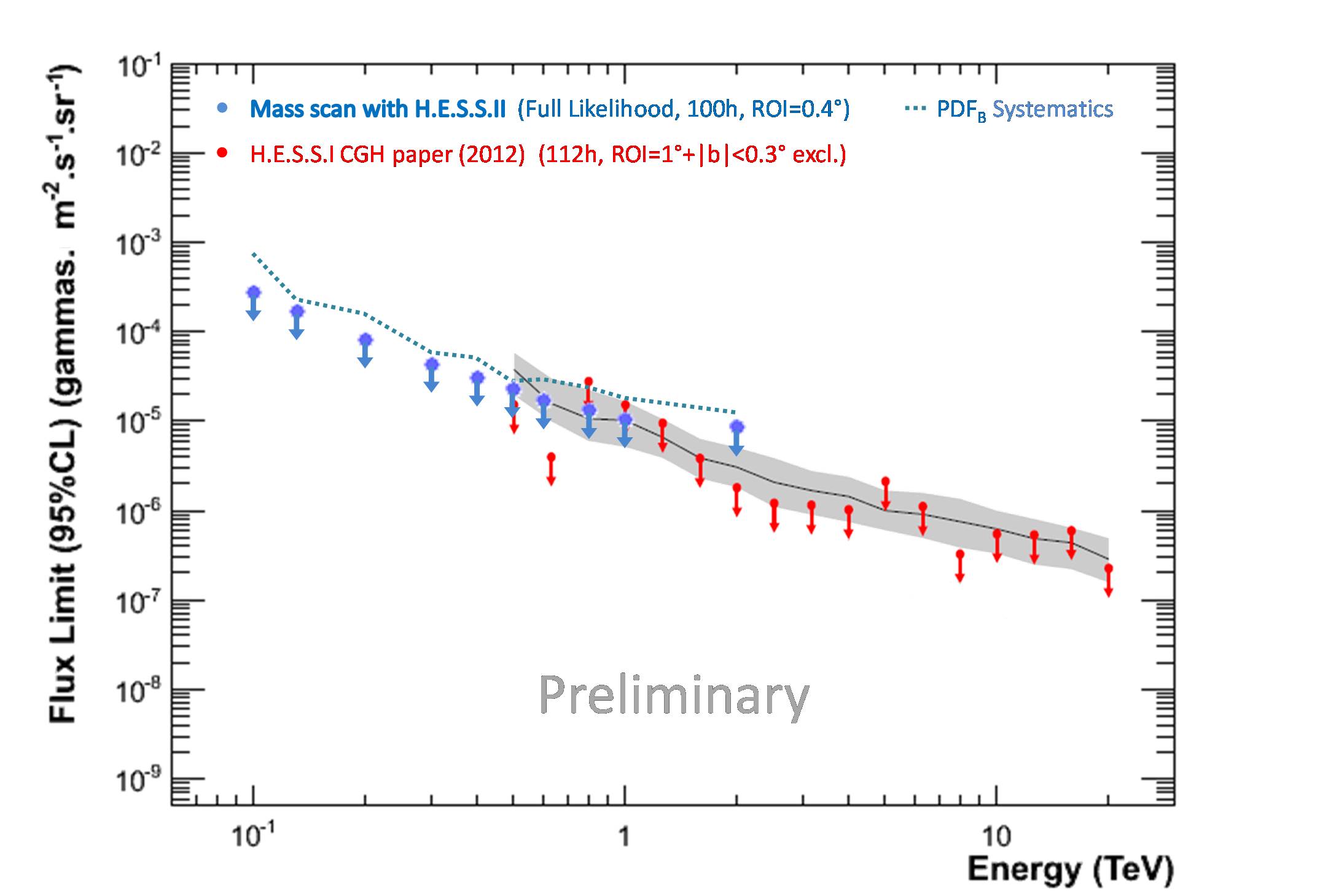}
\caption{Full likelihood results : Flux limits at 95\% CL for a line scan between 100 GeV and 2 TeV.  The flux is expressed in $\gamma\ \textrm{m}^{-2} \textrm{s}^{-1}$ and per steradian. The filled blue data points represent the computed MC limits with 100h of observation time and taking the median value of 500 simulations. Effects on the limits due to systematic uncertainties on the background PDF are represented by a dashed blue line. Former limits from H.E.S.S. I \cite{hess_line} are represented by red data points.}
\label{fig:mc_limits_flux}
\end{figure}

\begin{figure}[ht]
\centering
\includegraphics[width=1\textwidth]{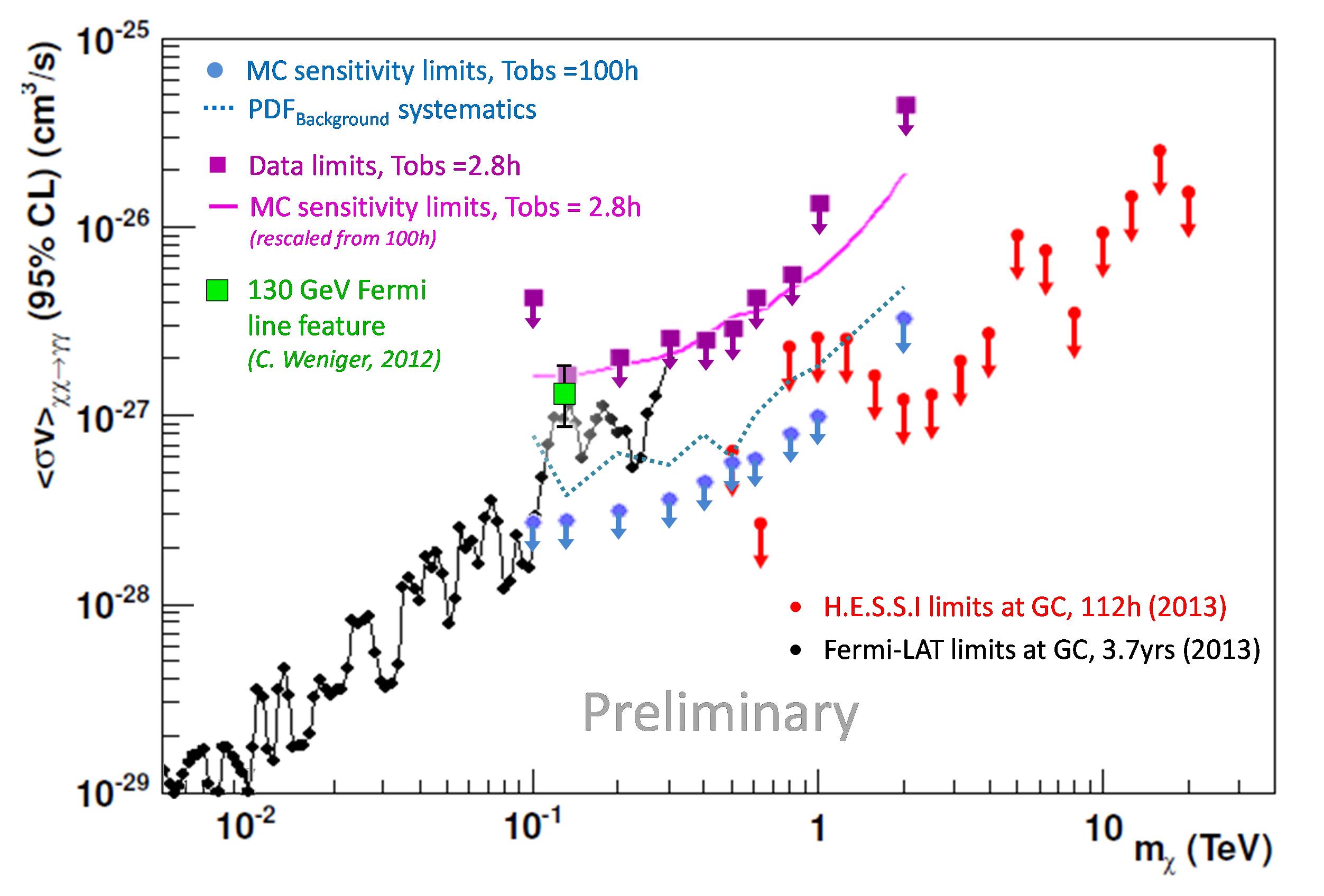}
\caption{Full likelihood results : $<\sigma v>$ limits at 95\% CL for the line scan between 100 GeV and 2 TeV, expressed in $\textrm{cm}^{3} \textrm{s}^{-1}$. The filled blue data points represent the computed MC limits assuming 100h of observation time and taking the median value of 500 simulations. Effects on the limits due to systematic uncertainties on the background PDF are represented by a dashed blue line. The data points in violet represent limits obtained from 2014 data sample
corresponding to 2.8 hours of observation time. For comparison, the expected MC limits for the same observation time (rescaled from the 100 hour values) are also shown with a violet solid line. Former limits from H.E.S.S. I \cite{hess_line} and Fermi-LAT \cite{bib:1} are represented by red and black data points, respectively. Finally the $<\sigma v>$ value corresponding the the 130 GeV line feature reported by C. Weniger \cite{bib:4} is shown in green.}
\label{fig:mc_limits_sigmav}
\end{figure}

The full likelihood method was performed with MC simulations to provide sensitivity estimates as a function of $E_{line}$ and to study the impact of systematic uncertainties. The numbers of measured background events in the ROI of $0.4^{\circ}$ and $PDF_{bckg}$ parametrization were derived from the control OFF-source regions around the Fermi hotspot. The form of the $PDF_{bckg}$ is shown in Figure \ref{fig:background_pdf}. The $PDF_{signal}$ was provided by a full MC simulation of a line signal at given energy $E_{line}$ smeared by the H.E.S.S. energy resolution. The full likelihood fits covered two pre-defined energy ranges from 80 GeV to 1 TeV and from 200 GeV to 3 TeV which allowed probing the sensitivity to line signals from 100 GeV to 500 GeV and from 500 GeV to 2 TeV, respectively, ensuring a large energy lever-arm in the fit in each case. Sets of 500 simulations were considered for each line energy of the scan assuming 100h of exposure, and led to 95\% CL upper limits first on $\eta$, then on the number of excess events $N_{\gamma}^{95\% CL}$. The resulting limit is defined as the median of the 500 simulations, as shown in the example in Figure \ref{fig:ngamma_limits_median}. \\
\\
\noindent
The corresponding limits on the flux ($\Phi$) and on the DM annihilation cross-section ($<\sigma v>$) are given by:
\begin{eqnarray}
\Phi^{95\% CL} = \frac{N_{\gamma}^{95\% CL}}{T_{OBS}} \times \frac{\int \limits_{E_{min}}^{E_{max}} \frac{dN}{dE_{\gamma}}(E_{\gamma})dE_{\gamma}}{\int \limits_{E_{min}}^{E_{max}} A_{eff}(E_{\gamma}) \frac{dN}{dE_{\gamma}}(E_{\gamma})dE_{\gamma}}\ \ \ \ \ , \ \ \ \ \ <\sigma v>^{95\% CL} = \frac{8 \pi m_{\chi}^{2}}{2\Phi_{astro}} \times \Phi^{95\% CL}
\end{eqnarray}
where  $T_{OBS}$ is the observation time, $A_{eff}$ and $\frac{dN}{dE}$ are respectively the effective area for gammas (in $m^{2}$) and the differential energy spectrum of the expected DM signal expressed as functions of the true energy, $m_{\chi}$ the DM particle mass, $[E_{min},E_{max}]$ the bounds of the energy range and $\Phi_{astro}$ the astrophysical factor. A DM Einasto profile \cite{einasto} with halo parameters given in \cite{hess_line} has been considered at the center of the ROI which leads to the value of $\Phi_{astro} = 2.46 \times 10^{21}\ \textrm{GeV}^{2} \textrm{cm}^{-5}$. For DM gamma-ray lines the differential energy spectrum is $\frac{dN}{dE_{\gamma}}(E_{\gamma}) = 2\delta(E_{\gamma}-m_{\chi})$ where the factor 2 results from the annihilation of DM particles into two gammas.
The impact of radial acceptance within the signal region (not considered in the calculation) has been assessed and is found to only affect the limits obtained at the few percent level. The signal region being sufficiently large there is no effect due to the point spread function (PSF). Also due to the large extension of the galactic DM halo a fraction of the expected DM signal leaks into the background regions, at the level of 25.1\% of the DM signal in the ROI. The presented $<\sigma v>$ limits account for that effect. Finally, nuisance parameters have been introduced in the full likelihood function to estimate the impact of systematic uncertainties in the limits calculation. The considered sources of errors are the IRFs, the background PDF shape and the diffuse emission component included in the background regions. 
\\ \\
Results on the flux and $<\sigma v>$ limits with MC data are presented with the blue data points in Figures \ref{fig:mc_limits_flux} and \ref{fig:mc_limits_sigmav}, respectively, and show the potential of the method for line signal detection. The flux is expressed per solid angle unit (with a division by $\Delta\Omega$) to be compared with the prior line analysis at the Galactic Center\cite{hess_line} performed with a larger ROI. As an example at 500 GeV, $\Phi^{95\% CL}/\Delta\Omega=2.3 \times 10^{-5}\ \gamma\ \textrm{m}^{-2} \textrm{s}^{-1} \textrm{sr}^{-1}$ and $<\sigma v>^{95\% CL}=5.6 \times 10^{-28}\ \textrm{cm}^{3} \textrm{s}^{-1}$. 
The data limits (violet) seen on Figure \ref{fig:mc_limits_sigmav} will be discussed later in section \ref{FermiHS}.
The background PDF shape has been identified to be the dominant source of systematic uncertainties in the limits calculation, represented with a dashed line.
The limits obtained with H.E.S.S. II are completing the former H.E.S.S.\textcolor{white}{i}I \cite{hess_line} and Fermi-LAT \cite{bib:1} scans and cover the gap in mass between 300 and 500 GeV. 

\section{Particular Case of the 130 GeV Line Feature}
\label{FermiHS}
\begin{figure}[t]
 \centering
 \includegraphics[width=0.8\textwidth]{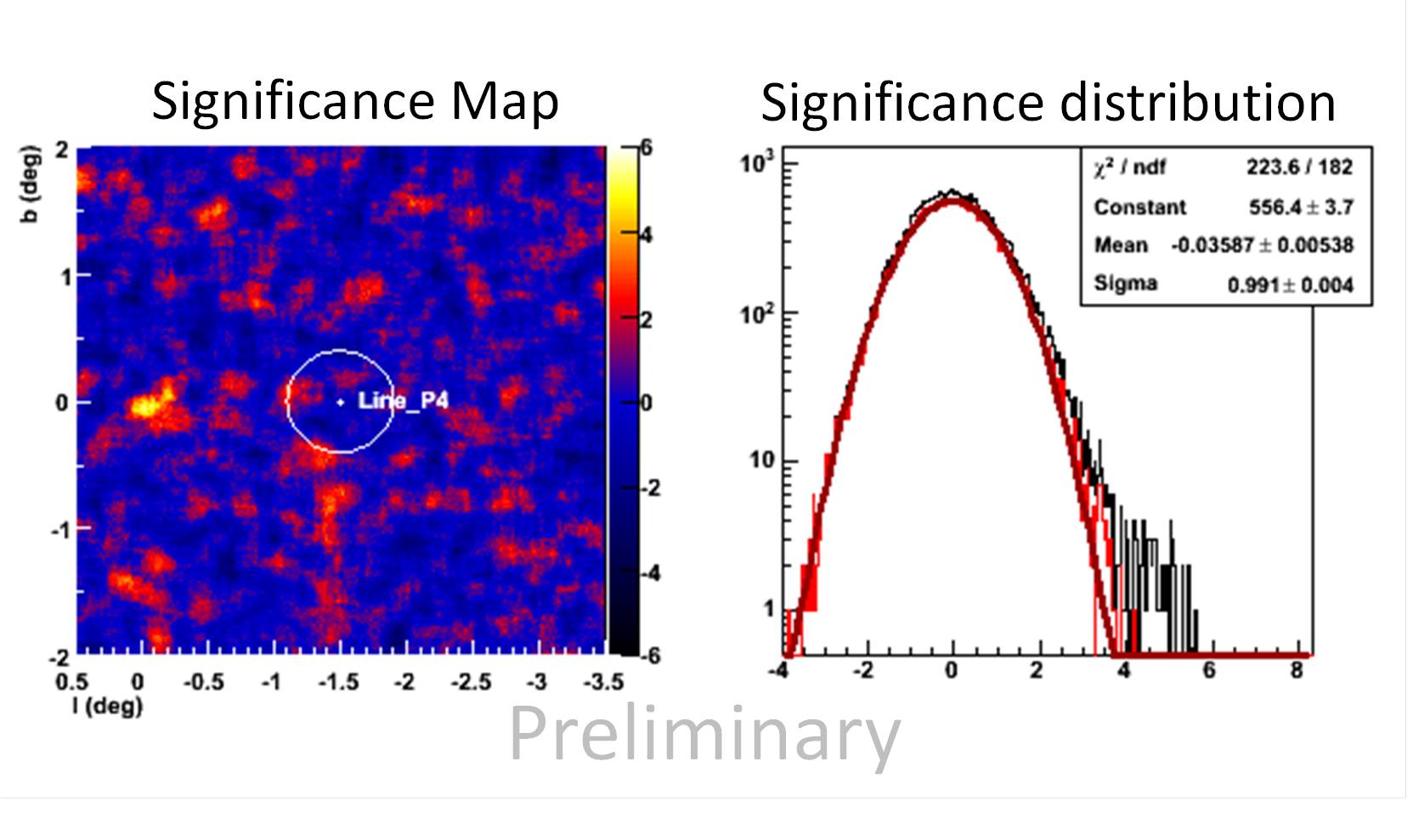}
 \caption{Significance map (left) and significance distribution (right) in the FOV. The ROI is represented with a white circle centred on the Fermi hotspot ($-1.5^{\circ}$, $0^{\circ}$). The known source HESS J1745-290 is clearly detected.}
 \label{fig:linescan_map}
\end{figure}

The centre of the 130 GeV excess as observed in the
Fermi-LAT data was found to be displaced with respect to the position of the Galactic
Centre by $-1.5^{\circ}$ galactic longitude, although with large
uncertainty \cite{su_fink}. Since the exact position of this excess
appears uncertain, H.E.S.S. observations were performed in a scan pattern along the Galactic plane,
with pointing positions ranging from $-2.3^{\circ}<l<0.5^{\circ}$ with a step size of
$0.7^{\circ}$ and $b=\pm0.8^{\circ}$ in galactic coordinates. \\
A total of $\sim$20h of data covering the Fermi hotspot position have been taken in spring and summer 2014 of which only 2.8 hours have been analysed till now. At least 4 telescopes were requested for ensuring good reconstruction of the gamma ray events, including the large CT5 telescope and considering observations at low zenith angles ($< 20^{\circ}$) to guarantee the lowest possible energy threshold. Data quality checks have been performed by checking the global run and the individual telescope status. Cuts have been applied on the telescope trigger rates, trigger rate stability and broken pixels fraction in the camera in order to remove bad quality runs from the analysis.
After data quality selections, several criteria on photon$-$hadron discrimination were applied to the reconstructed event samples in the signal and background regions leading to an effective energy threshold of $E_{\textrm{th}}$ $\sim$ 80 GeV. 
\\ \\
Then significance maps were reconstructed for the 2.8h dataset as shown in Figure \ref{fig:linescan_map}. The distribution of significances derived from background fluctuations follow a Gaussian distribution as expected, with a significant excess in the Field-of-View at the position of HESS J1745-290. No significant excess ($< 1 \sigma$) is found in the $0.4^{\circ}$ ROI at the best-fit position of the Fermi hotspot $(l,b)=(-1.5^{\circ}, 0^{\circ})$. 

\noindent
Consequently, 95\% CL upper limits on the flux and $<\sigma v>$ have been derived from the full likelihood fit on the data sample following exactly the same procedure as for the MC simulations discussed in section \ref{MC} and considering the same DM Einasto profile \cite{hess_line}. 
Limits for the complete scan between 100 GeV and 2 TeV are shown in Figure \ref{fig:mc_limits_sigmav} (violet data points) and are compared with the expected limits from the MC simulations, rescaled to 2.8h of observation time. Very good compatibility between MC predictions and the data analysis is observed. 
The related data/MC values at 130 GeV are presented in Table \ref{tab:table1}.
\begin{table}
\begin{center}
\begin{tabular}{|l|c|c|}
\hline
 & $\Phi^{95\% CL}/\Delta\Omega$ & $<\sigma v>^{95\% CL}$ \\
 & $10^{-4}\ \gamma\ \textrm{m}^{-2} \textrm{s}^{-1} \textrm{sr}^{-1}$  &  $10^{-27}\ \textrm{cm}^{3} \textrm{s}^{-1}$ \\
\hline
Data & $9.96 $ & $ 1.64 $ \\
\hline
MC & $9.95 $ & $ 1.64 $ \\
\hline
\end{tabular}
\end{center}
\caption{95\% CL limits on the flux (per solid angle unit) and $<\sigma v>$ from the likelihood fit for the detection of the 130 GeV line. The MC values are coming from the MC line scan presented in section \protect\ref{MC}, rescaled to 2.8h of observation time.}
\label{tab:table1}
\end{table}
The limit on $<\sigma v>$ at 130 GeV exceeds the best fit value of \cite{bib:4} value by a factor of 1.3 with $\sim$3 hours of observation time. Only a substantial increase in observation time would lead to more stringent limits on the cross-section found by C. Weniger \cite{bib:4}. \\
A cross-check study was performed on the source data sample with different calibration and reconstruction methods. The preliminary results confirm the conclusion of an insignificant excess at 130 GeV as obtained in the analysis described above.

\section{Conclusions}
The present analysis of the H.E.S.S. data including CT5 in the setup suggests an excellent potential for line reconstruction between $\sim$100 GeV and and a few TeV, considering 100h of observation time. The study was performed with background estimated in the Galactic Centre region and yielded limits on the gamma-ray flux and on $<\sigma v>$ which efficiently fill the gap between results from Fermi-LAT and the first phase of H.E.S.S. 
On the other side, the analysis of a limited set of data corresponding to $\sim$3 hours of observation time shows no hint of an excess at 130 GeV at $95\%$ CL in the region of the Fermi hotspot $(l,b)=(-1.5^{\circ}, 0^{\circ})$.

\end{document}